\documentclass[english,aip,apl,reprint]{revtex4-1}
\usepackage[T1]{fontenc}
\usepackage[latin9]{inputenc}
\usepackage{float}
\usepackage{amstext}
\usepackage{graphicx}
\usepackage{wasysym}
\usepackage{esint}

\makeatletter
\usepackage{epstopdf}
\usepackage{hyperref}
\hypersetup{
  linkcolor  = blue,
  citecolor  =blue,
  urlcolor   = blue,
  colorlinks = true,
}

\makeatother

\usepackage{babel}
\begin{document}

\title{Microchannel plate cross-talk mitigation for spatial autocorrelation
measurements}

\author{Micha\l{} Lipka}
\email{michal.lipka@fuw.edu.pl}

\affiliation{Institute of Experimental Physics, Faculty of Physics, University
of Warsaw, Pasteura 5, 02-093 Warsaw, Poland}

\author{Micha\l{} Parniak}

\affiliation{Institute of Experimental Physics, Faculty of Physics, University
of Warsaw, Pasteura 5, 02-093 Warsaw, Poland}

\author{Wojciech Wasilewski}

\affiliation{Institute of Experimental Physics, Faculty of Physics, University
of Warsaw, Pasteura 5, 02-093 Warsaw, Poland}
\begin{abstract}
Microchannel plates (MCP) are the basis for many spatially-resolved
single-particle detectors such as ICCD or I-sCMOS cameras employing
image intensifiers (II), MCPs with delay-line anodes for the detection
of cold gas particles or Cherenkov radiation detectors. However, the
spatial characterization provided by an MCP is severely limited by
cross-talk between its microchannels, rendering MCP and II ill-suited
for autocorrelation measurements. Here we present a cross-talk subtraction
method experimentally exemplified for an I-sCMOS based measurement
of pseudo-thermal light second-order intensity autocorrelation function
at the single-photon level. The method merely requires a dark counts
measurement for calibration. A reference cross-correlation measurement
certifies the cross-talk subtraction. While remaining universal for
MCP applications, the presented cross-talk subtraction in particular
simplifies quantum optical setups. With the possibility of autocorrelation
measurement the signal needs no longer to be divided into two camera
regions for a cross-correlation measurement, reducing the experimental
setup complexity and increasing at least twofold the simultaneously
employable camera sensor region. 

\end{abstract}
\maketitle
Single-excitation level spatially-resolved detectors find profound
applications in research ranging from optics, atomic physics \citep{Jeltes2007,Schellekens648}
or high-energy physics \citep{Gys2015} to \emph{in-vivo} imaging
in medicine \citep{Hui2014} and biology \citep{Pian2017}, or nanoscale
material science\citep{Rezaei2018}. In particular, single-photon
sensitive cameras constitute a cornerstone of development in quantum
optics, information processing \citep{Picken2017}, computing \citep{Tasca2011}
and communications\citep{Zhou2016,Tentrup2017}; enabling superresolution
imaging \citep{Parniak2018,Schwartz2013} or localization \citep{Israel2017};
bringing to life versatile, multi-mode quantum memories \citep{Parniak2017,Leszczynski2018}
and fostering better comprehension of nonclassical light \citep{Dabrowski2017,Haderka2005,Reichert2017,Aspden2013,Edgar2012,Qi2018,Just2014}.
Some of those have been achieved with charge coupled device (CCD)
cameras with an image intensifier (ICCD) or electron multiplying CCD
(EMCCD) cameras; however, their applicability is severely limited
by either a low read-out speed or high read-out noise. Although these
obstacles have been overcome to a great extent by the recently emerged
scientific complementary metal oxide semiconductor (sCMOS) cameras
coupled (I-sCMOS) with an image intensifier (II) \citep{Picken2017,chrapphd},
the sole II introduces artificial, strongly correlated photon counts
\citep{Korpar2008}. This deleterious effect in II can be attributed
to cross-talk between microchannels of a microchannel plate (MCP).
Cross-talk can be detrimental in the light autocorrelation measurements,
veiling the true light correlations. Inability to perform autocorrelation
measurements either severely limits the thoroughness of spatial light
characterization or enforces a cross-correlation measurement instead,
with the optical signal divided into two separate regions of the camera.
In such a case, not only the effective camera frame size is reduced
by at least a half, but more importantly, often inconvenient complications
in the experimental setup emerge, possibly degrading the measurement
quality. Furthermore, with photons registered on two separate regions
a cross-correlation measurement neglects a half of photon-pair events.
\begin{figure}[H]
\centering \includegraphics[width=8.5cm]{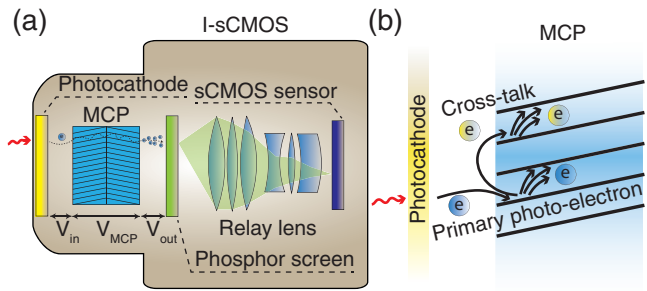} 

\caption{(a) Schematic of the I-sCMOS camera operating a the single-photon
level. (b) MCP cross-talk mechanism. Secondary emission electrons
may be scattered backwards to leave the MCP and re-enter a distant
microchannel. }

\label{fig:crsmech} 
\end{figure}

Here we present a straightforward method of cross-talk subtraction
relaying on a simple, dark counts calibration measurement. The method
enables reconstruction of the second-order light intensity autocorrelation
function $g^{(2)}$ for a priori unknown light. We exemplify the cross-talk
subtraction by employing an I-sCMOS camera to measure pseudo-thermal
light $g^{(2)}$. A separate, reference cross-correlation measurement
certifies the performance of our method. While here we discuss the
results for light correlation measurements, the cross-talk subtraction
method remains valid whenever an MCP is employed for spatially resolved
detection. Our method supplements a broad range of quantum efficiency
and spectral calibration techniques \citep{Perina2014,Haderka2014,Qi2016,Avella2016,Chrapkiewicz2014}
for spatially resolved single-photon detectors as well as cross-talk
calibration techniques \citep{Kalashnikov2012,Gallego2013,Rosado2015}
concerned with photon counting on photon number resolving detectors
providing no spatial light characterization.
\begin{figure}[H]
\centering \includegraphics[width=8.5cm]{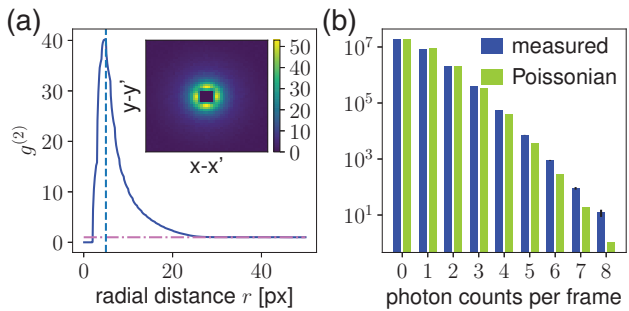} 

\caption{(a) \emph{inset:} Dark counts measurement reveals cross-talk induced
correlations in the map of the second order intensity autocorrelation
$g^{(2)}(x-x',y-y')$ in the difference coordinates.\emph{ main plot:}
Dark counts $g^{(2)}(x-x',y-y')$ averaged over the polar angle, yielding
a radial map of autocorrelation $g^{(2)}(r)$. (b) Dark counts full-frame
counts number statistics in the presence of cross-talk (\emph{measured})
becomes super-Poissonian (c.f. \emph{Poissonian}). }

\label{fig:crsg2} 
\end{figure}

We employed a custom I-sCMOS camera constructed of commercially available
components. Phosphor screen (P43 phosphor) of the employed image intensifier
(Hammamatsu V7090D-71-G232) is imaged by a relay lens (Stanford Computer
Optics, $f/\#=1.1$, magnification $M=-0.44$) onto an $5.5\;{\rm Mpx}$,
$2560\times2160\;{\rm px}$ sCMOS sensor of Andor Zyla 5.5 camera
(effective pixel pitch ${\rm px}\approx15\;\mu{\rm m}$ accounting
for $M$) operated in \emph{rolling shutter }and \emph{overlap} modes.
As illustrated in figure \ref{fig:crsmech} (a), during the camera
operation at the single-photon level, an incident photon (a prior)
strikes the photocathode releasing an electron which is accelerated
by a gated potential $V_{{\rm in}}=-50\dots200\;{\rm V}$ towards
the MCP. Upon colliding with the MCP wall, secondary emission electrons
are released beginning the avalanche gain process. Electrons are further
accelerated in a potential across the MCP $V_{{\rm MCP}}=250\dots1750\;{\rm V}$
and an output potential $V_{{\rm out}}=6\;{\rm kV}$ between the MCP
and the phosphor screen. The electrons leaving the MCP (overall gain
ca. $10^{5}$) strike the phosphor screen producing bright flashes
imaged onto the sCMOS sensor. A data reduction, real-time procedure
(\emph{photonfinder)} localizes the flashes maxima and performs an
image to flash (photon) position conversion.

Some of the electrons released in a microchannel may be scattered
backwards and leave the MCP, as illustrated in figure \ref{fig:crsmech}
(b). Such electrons may re-enter a distant microchannel, thereby starting
an another electron avalanche. The involved time scales are well below
the MCP gate impulse duration and the I-sCMOS inter-frame period;
therefore, the additional avalanche (a cross-talk event) is registered
in the same camera frame as its prior, yielding strong, artificial
correlations decaying on a scale of ca. $20\;{\rm px}$ in the difference
between the positions of the prior and the cross-talk event. To quantify
the cross-talk, let us consider a 4-dimensional second order photon
counts autocorrelation function $g^{(2)}(x,x';\;y,y')$ given by:

\begin{equation}
g^{(2)}(x,x';\;y,y')=\frac{\langle n(x,y)n(x',y')\rangle}{\langle n(x,y)\rangle\langle n(x',y')\rangle},\label{eq:g2gen}
\end{equation}
where $n(x,y)\in\{0,1\}$ corresponds to the number of photon counts
at a camera pixel with coordinates $(x,y)$, during on-off operation
in the single-photon regime. Here $x\in[1,L_{x}],\;y\in[1,L_{y}]$,
where $L_{x}\times L_{y}$ is the size of the autocorrelated region
in px. We shall now recast equation \ref{eq:g2gen} in the sum $s_{x}=x+x',\;s_{y}=y+y'$
and difference $d_{x}=x-x',\;d_{y}=y-y'$ coordinates by substituting
$x\rightarrow(s_{x}+d_{x})/2,\;x'\rightarrow(s_{x}-d_{x})/2;\:y\rightarrow(s_{y}+d_{y}),\:y'\rightarrow(s_{y}-d_{y})/2$
to obtain $g^{(2)}(d_{x},s_{x};\;d_{y},s_{y})$. Note that $|d_{x}|\in[0,L_{x}-1]$,
$s_{x}\in[2+|d_{x}|,2L_{x}-|d_{x}|]$ and $s_{x}$ takes on discrete
values with an increment of 2. Analogous relations are valid for the
$y$ dimension. By averaging $g^{(2)}(d_{x},s_{x};\;d_{y},s_{y})$
over the sum coordinates $s_{x},s_{y}$, a 2-dimensional autocorrelation
map $g^{(2)}(d_{x},d_{y})$ is obtained, as given by:
\begin{equation}
g^{(2)}(d_{x},d_{y})={\cal N}\sum_{j,l}g^{(2)}(d_{x},2j+|dx|;\;d_{y},2l+|dy|),\label{eq:g2diff}
\end{equation}
with the summation over $j\in[1,L_{x}-|d_{x}|]$ and $l\in[1,L_{y}-|d_{y}|]$
and with ${\cal N=\mathnormal{1/(}{\cal N}}_{x}{\cal N}_{y})$, where
${\cal N}_{x}=L_{x}-|d_{x}|$ (${\cal N}_{y}=L_{y}-|d_{y}|$) corresponds
to the number of distinct $s_{x}$ ($s_{y}$) values on the anti-diagonal
$x-x'=d_{x}$ ($y-y'=d_{y}$) contained within the $L_{x}\times L_{x}$
($L_{y}\times L_{y})$ region. The cross-talk affected photon counts
autocorrelation $g^{(2)}(d_{x},d_{y})$ has been illustrated in the
inset of figure \ref{fig:crsg2} (a) for $N_{\text{frm}}=3\times10^{7}$
frames ($L_{x}\times L_{y}=225\times435\;{\rm px}$) of dark counts,
registered at a framerate of $300\;{\rm fps}$ with an MCP gate time
of $4\;\mu{\rm s}$, yielding an average of $\bar{N}=0.47$ photon
counts per frame. Even though for Poissonian dark counts a lack of
correlation $g^{(2)}=1$ is expected, $g^{(2)}$ reaches up to ca.
40 due to the cross-talk induced correlations. Note that except in
the closest vicinity of the primary photon $(x-x',\;y-y'<5\;{\rm px})$,
the cross-talk is isotropic. Therefore, we shall depict $g^{(2)}(d_{x},d_{y})$
in the polar coordinates, averaging over the polar angle to obtain
$g^{(2)}(r)$, as illustrated in figure \ref{fig:crsg2} (a). The
close-range anisotropy can be accounted for by the inner workings
of the image-position conversion of the \emph{photonfinder}, unable
to localize photons registered closer than $3\;{\rm px}$ apart on
a single camera frame. The presence of cross-talk can be also observed
in the full frame photon counts statistics being super-Poissonian,
as depicted in figure \ref{fig:crsg2} (b). Note that even though
$g^{(2)}(x,x';\;y,y')$ is truncated at the camera frame border, $g^{(2)}(d_{x},d_{y})$
gives a measure of correlated photon counts positions difference and
therefore remains to some extent immune to deterioration due to edge
effects as long as the distance between the furthest observed correlated
photon counts remains well below linear dimensions of the camera frame
i.e. several full expected ranges of $d_{x},d_{y}$ can be observed
in the $L_{x}\times L_{y}$ frame region. Due to a vast area of the
image intensifier of ca. $10^{6}\;\text{px}$ this condition can be
fulfilled for light measurements by employing a proper magnification.
Nevertheless, for larger $|d_{x}|$ ($|d_{y}|$) averaging in equation
\ref{eq:g2diff} is performed over smaller photon count statistics.

For the purpose of cross-talk subtraction, let us now consider separately
an unnormalized 4-dimensional map of coincidences $\langle n(x,y)n(x',y')\rangle$
constituting the $g^{(2)}(x,x';\:y,y')$ numerator, as given by equation
\ref{eq:g2gen}, and the normalization factor of $\langle n(x,y)\rangle\langle n(x',y')\rangle$
present in the denominator. By performing a coordinate transformation
to the sum $s_{x},s_{y}$ and the difference $d_{x},d_{y}$ coordinates
and integrating out the former, we obtain 2-dimensional maps of the
coincidences (from $\langle n(x,y)n(x',y')\rangle$) and the normalization
factor (from $\langle n(x,y)\rangle\langle n(x',y')\rangle$). The
calculation is in a perfect analogy to equation \ref{eq:g2diff} for
$g^{(2)}$. Obtained maps can be now transformed to polar coordinates
and after multiplying by the number of frames $N_{\text{frm}}$ and
averaging over the polar angle may be again combined to yield $g^{(2)}(r)=c_{c}(r)/c_{a}(r)$.
Here $c_{c}(r)$ is obtained from $\langle n(x,y)n(x',y')\rangle N_{\text{frm}}$
and corresponds to the total number of registered coincidences, whereas
$c_{a}(r)$ originates from $\langle n(x,y)\rangle\langle n(x',y')\rangle N_{\text{frm}}$
and gives the total number of accidental coincidences i.e. the expected
number of coincidences if the photon counts were spatially uncorrelated.
For ideal Poissonian light $g^{(2)}(r)=1$ therefore $c_{c}(r)=c_{a}(r)$.
The difference $c_{r}(r)=c_{c}(r)-c_{a}(r)$ corresponds to the transformed
photon counts covariance $N_{\text{frm}}{\rm Cov}[n(x,y),n(x',y')]=N_{\text{frm}}[\langle n(x,y)n(x',y')\rangle-\langle n(x,y)\rangle\langle n(x',y')\rangle]$.
For Poissonian dark counts measurement a non-vanishing $c_{r}(r)$
indicates and quantifies coincidences originating from cross-talk
events. Therefore, $c_{r}(r)$ calculated from a dark counts measurement
provides information on the radial cross-talk probability distribution
and the total number of cross-talk induced events. 

To arrive at a method of cross-talk subtraction, let us now assume
that each real, registered photon (a prior) has a small $p_{c}\ll1$
probability of inducing exactly one registered cross-talk event. Under
this assumption, even with many photons per frame, each induced cross-talk
event only contributes a single spatially localized coincidence together
with its prior. Coincidences between cross-talk events originating
from different priors and between cross-talk events and photon counts
not being their priors are spatially random, therefore contribute
equally to $c_{c}(r)$ and $c_{a}(r)$ leaving the cross-talk term
$c_{r}(r)$ unchanged. Therefore, it is justified to assume a linear
scaling between the total number of cross-talk coincidences $\int c_{r}$
and the total number of photon counts $N_{p}$ across all $N_{\mathrm{frm}}$
frames, i.e. $\int c_{r}=\alpha N_{\mathrm{frm}}\bar{N}=\alpha N_{p}$.
If $\bar{n}$ real photons are on average registered per frame, then
$\bar{N}=(1+p_{c})\bar{n}$ and the total number of cross-talk events
can be expressed as $\int c_{r}=p_{c}\bar{n}=\bar{N}p_{c}/(1+p_{c})$
, therefore $\alpha=p_{c}/(1+p_{c})$ with $p_{c}$ interpreted as
a probability of a registered cross-talk event per a real, registered
photon. Note that if we forgo the assumption of $p_{c}\ll1$, each
prior photon could induce several cross-talk events. In such a case,
not only coincidences between the prior and each of its cross-talk
events would contribute to $\int c_{r}$ (linear scaling with $N_{p}$)
but also so would coincidences among the cross-talk events originating
from the same prior (super-linear scaling with $N_{p}$).

Using the linear scaling between $N_{p}$ and $\int c_{r}$, once
$c_{r,{\rm cal}}(r)$ has been measured in a calibration \emph{cal}
dark counts measurement with a total of $N_{p,{\rm cal}}$ registered
photon counts, any autocorrelation measurement (measurement under
correction) \emph{muc} can be rectified by removing a portion $c_{r,\text{cal}}(r)N_{\text{\ensuremath{\mathnormal{p}},muc}}/N_{\text{\ensuremath{\mathnormal{p}},cal}}$
of all measured coincidences $c_{\text{\ensuremath{\mathnormal{c}},muc}}(r)$.
In this way a corrected $g_{\text{corr}}^{(2)}(r)$ is obtained, as
given by: 
\begin{equation}
g_{\text{corr}}^{(2)}(r)=g_{\text{raw}}^{(2)}(r)-\frac{c_{r,\text{cal}}(r)N_{\text{\ensuremath{\mathnormal{p}},muc}}}{c_{a,\text{muc}}(r)N_{\text{\ensuremath{\mathnormal{p}},cal}}},\label{eq:dutcross-1}
\end{equation}
where $g_{\text{raw}}^{(2)}(r)=c_{c,\text{muc}}(r)/c_{a,\text{muc}}(r)$
corresponds to the measured \emph{muc} autocorrelation function before
correction. Note that such a method of cross-talk subtraction remains
cross-talk model independent, relying solely on the assumption of
linear scaling between the number of cross-talk events $\int c_{r}$
and photon counts $N_{p}$ and the assumption of cross-talk spatial
probability distribution being isotropic and independent of the true
light correlations. Under an additional assumption of equal spatial
intensity distributions for \emph{cal }and \emph{muc} measurements
$c_{a}(r)\propto N_{p}^{2}/N_{\text{frm}}$ with the same radial dependence
for both measurements. In such a case equation \ref{eq:dutcross-1}
can be simplified to rely only on $g^{(2)}(r)$ functions and experimental
parameters $g_{\text{corr}}^{(2)}(r)=g_{\text{raw}}^{(2)}(r)-[g_{\text{cal}}^{(2)}-1]N_{p,\text{cal}}N_{\text{frm,muc}}/(N_{p,\text{muc}}N_{\text{frm,cal}})$,
where we have used $c_{r,\text{cal}}(r)=c_{a,\text{cal}}(r)[g_{\text{cal}}^{(2)}-1]$.
\begin{figure}[H]
\centering \includegraphics[width=8.5cm]{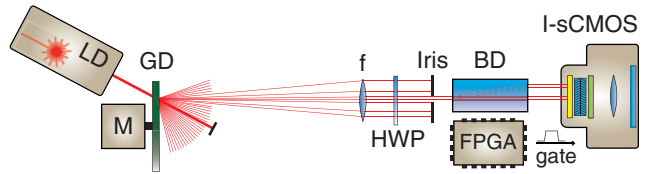} 

\caption{Experimental setup. Psuedo-thermal light generated with a rotating
ground glass diffuser (GD) is far field imaged onto the I-sCMOS camera
sensor. }

\label{fig:expsetup} 
\end{figure}

To certify the subtraction method we have performed an autocorrelation
measurement of pseudo-thermal light obtained in a standard \citep{martienssen64}
setup, with a Gaussian beam of a diode laser (780 nm, central wavelength
$\lambda$) scattered on a rotating ground glass diffuser (GD, grit
600), as illustrated in figure \ref{fig:expsetup}. GD is far field
imaged  onto the I-sCMOS sensor using a single lens of a focal length
$f=250\;{\rm mm}$. A beam displacer (BD) splits the pseudo-thermal
light onto two separate regions on the camera. While an iris placed
near the far-field of the GD ensures the regions do not overlap, an
equal power division ration between the regions is obtained with a
half-wave plate (HWP). In this Handbury Brown-Twiss type \citep{hbt56}
setup both autocorrelation (in one of the regions) and cross-correlation
(between the regions) can be measured. Note that both measurements
share nearly the same experimental imperfections. Therefore, cross-correlation
being immune to the cross-talk induced coincidences can serve as a
reference for the quality of cross-talk subtraction in the autocorrelation. 

For the pseudo-thermal light, we expect the second order intensity
correlation function $g^{(2)}(r)$ to be given by:
\begin{equation}
g^{(2)}(r)=1+\frac{1}{{\cal M}}\exp(-\frac{r^{2}}{2\sigma^{2}}),\label{eq:g2therm}
\end{equation}
where ${\cal M\geq}1$ denotes the effective number of pseudo-thermal
modes per one pixel and $\sigma=f\lambda/(w_{0}\sqrt{\pi})$ with
$w_{0}$ being the Gaussian RMS width of the incident beam intensity
$I_{{\rm GD}}(r)=\exp[-r^{2}/(2w_{0}^{2})]$ at GD. By recollimating
the beam and adjusting the laser power we could shape the correlation
width $\sigma$ and alter the average number of photon counts per
frame $\bar{N}$, respectively.

\begin{figure*}
\centering \includegraphics[width=17cm]{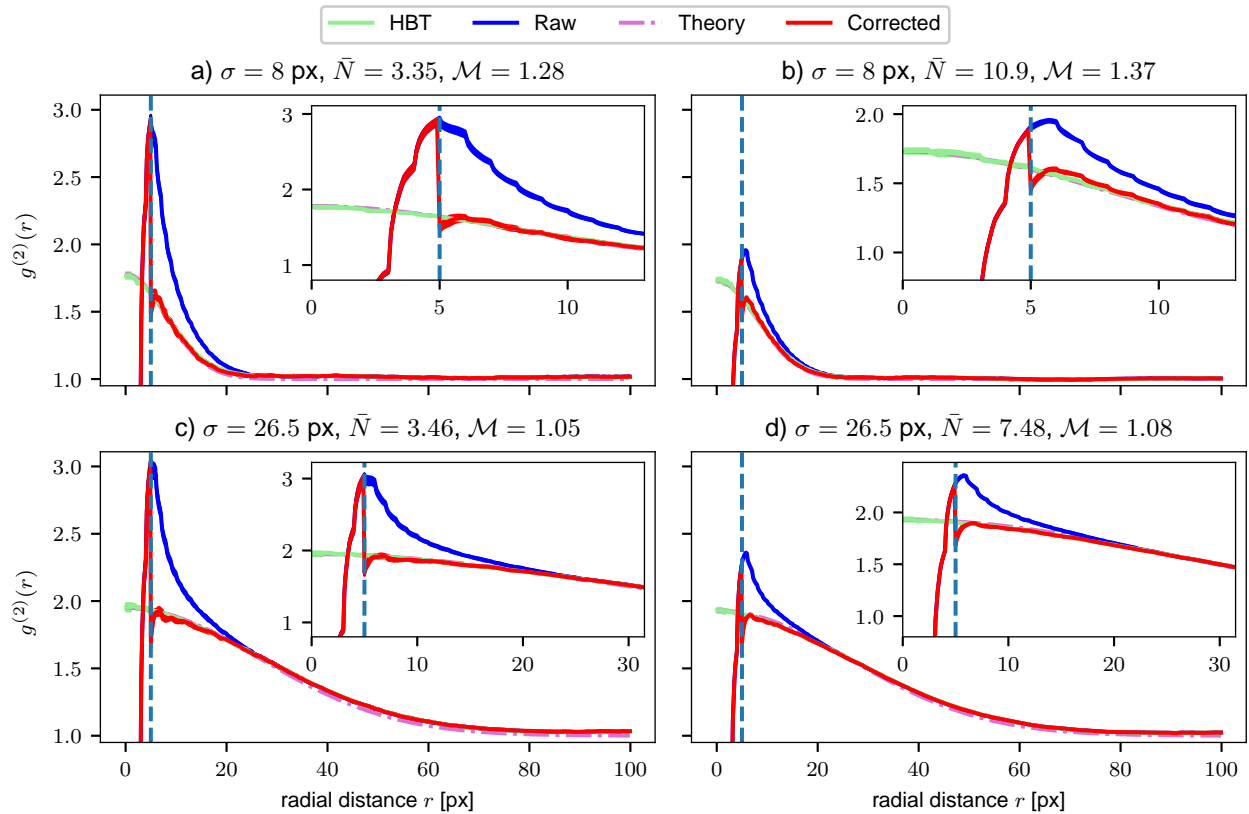} 

\caption{Radial dependence of the second order correlation function $g^{(2)}(r)$
for the pseudo-thermal light. From the raw autocorrelation measurement
(blue line), cross-talk induced $g^{(2)}$ is subtracted, yielding
the corrected autocorrelation (red line) in agreement with the reference
cross-correlation (green line) and the theoretical prediction (dot-dashed
line). Vertical line indicates the range $r\apprle5\;{\rm px}$ of
the artificially introduced reduction in $g^{(2)}$ due to the real-time
data reduction procedure.   }

\label{fig:g2crossvsauto} 
\end{figure*}

Figure \ref{fig:g2crossvsauto} depicts a sound agreement of the autocorrelation
measurement (\emph{Raw}) after cross-talk subtraction (\emph{Corrected})
with the reference cross-correlation (\emph{HBT}) and the theoretical
prediction (\emph{Theory}). The effective number of pseudo-thermal
modes ${\cal M}$ was adjusted to match the theoretical $g^{(2)}(0)$
to the cross-correlation measurement. Note that in the low optical
power regime (a), (c) the cross-talk induced autocorrelation dominates
over the true light $g^{(2)}$. Cross-talk subtraction reconstructs
correct $g^{(2)}(r)$ for $r\apprge5\;{\rm px}$, even with a pseudo-thermal
mode size (coherence length $\sigma\sqrt{2\pi}\approx2.5\sigma$)
comparable to the cross-talk range of ca. 20 px (a), (b). The uncertainty
of one standard deviation, indicated by the linewidths, has been calculated
by splitting the data into 10 sub-measurements with $1/10$ of the
total number of frames each and repeating the calibration - subtraction
procedure for each sub-measurement. 

For the cross-talk calibration $1.5\times10^{7}$ frames ($650\times200\;{\rm px}$)
of dark counts have been gathered with an average of $\bar{N}_{{\rm cal}}=1.2\times10^{-1}$
counts per frame, while the successive measurements depicted in figure
\ref{fig:g2crossvsauto} contained (a) $10^{7}$, (b) $10^{6}$, (c)
$10^{7}$, (d) $5\times10^{6}$ frames. The cross-correlated regions
were $180\times180\;{\rm px}$ each. Autocorrelation was calculated
for one of the regions. Each measurement of figure \ref{fig:g2crossvsauto}
has been annotated with the average number of photon counts $\bar{N}$
per the autocorrelated region.

The cross-talk proportionality constant $\alpha$ was calibrated at
$\alpha=(2.07\pm0.01)\times10^{-2}$ corresponding to the cross-talk
probability per photon of $p_{c}=(2.11\pm0.01)\times10^{-2}$. Even
though the cross-talk probability is on the order of $1\%$, in the
low optical power regime the cross-talk induced coincidences remain
dominant e.g. in the measurement (a) non-accidental coincidences ($\int c_{c}-c_{a}=(7.90\pm0.01)\times10^{4})$
contained $26\%$ of cross-talk induced events.

The central part of correlation function $g^{(2)}(r<5\;{\rm px})$
constituting an area of ca. $80\;{\rm px}$ was inaccessible in an
autocorrelation measurement, as illustrated in figure \ref{fig:g2crossvsauto};
however, with an employable image intensifier area of ca. $10^{6}\;{\rm px}$
the optical magnification can be adjusted to make the central $80\ {\rm px}$
area negligible. Furthermore, the inability to access this part of
the autocorrelation function follows only from the inner workings
of the real-time software - the \emph{photonfinder }employed to discriminate
photon counts in the raw images from the I-sCMOS camera. In principle,
with each pixel as a binary on/off detector the inaccessible central
area is of ca. $1\;{\rm px}$. 

In conclusion, we have presented a method of microchannel plate cross-talk
subtraction, enabling second-order autocorrelation measurements on
the single-excitation level. While the method remains universal as
long as an MCP is employed, we have experimentally provided an exemplary
application for the measurement of the second-order intensity autocorrelation
of pseudo-thermal light on an I-sCMOS camera with an image intensifier,
operated in the single-photon regime. Enabling the autocorrelation
measurements, our method greatly simplifies experimental setups where
the correlated signal would be previously divided into two MCP regions
to measure cross-talk insensitive cross-correlation. In the same way,
the MCP region that can be employed for correlation measurements is
increased at least twofold. Furthermore, removal of abundant signal
splitting mitigates the experimental imperfections, promising an enhanced
spatial characterization which is for the case of single-photon light
a cornerstone of modern quantum information processing technologies.

This work has been supported by the National Science Centre (Poland)
Projects No. 2017/25/N/ST2/01163, 2016/21/B/ST2/02559. We acknowledge
a generous support of K. Banaszek.

\bibliographystyle{apsrev4-1}
\bibliography{corr_meas_bib}

\end{document}